# Correlation of optical conductivity and ARPES spectra of strong-coupling large polarons and its display in cuprates

A.E.Myasnikova, E.N. Myasnikov

South Federal University, 344090 Rostov-on-Don, Russia

Common approach is used to calculate band due to strong-coupling large polaron (SCLP) photodissociation in ARPES and in optical conductivity (OC) spectra. It is based on using the coherent-states representation for the phonon field in SCLP. The calculated positions of both band maximums are universal functions of one parameter – the SCLP binding energy $E_p$: ARPES band maximum lies at binding energy about $3.2E_p$; the OC band maximum is at the photon energy about $4.2E_p$. The half-widths of the bands are mainly determined by $E_p$ and slightly depend on electron-phonon coupling constant α, for α =6-8 the ARPES band half-width is $1.7$-$1.3E_p$ and the OC band half-width is $2.8$-$2.2E_p$. Using these results one can predict approximate position of ARPES band maximum and half-width from the maximum of mid-IR OC band and vice versa. Comparison of the results with experiments leads to a conclusion that underdoped cuprates contain SCLPs with $E_p$=0.1-0.2 eV that is in good conformity with the medium parameters in cuprates. The values of the polaron binding energy determined from experimental ARPES and OC spectra of the same material are in good conformity too: the difference between them is within 10 %.

## 1. Introduction

ARPES (energy dispersion curves) as well as mid-IR optical conductivity (OC) spectra of high-temperature superconductors and other complex oxides demonstrate broad bands [1-17] naturally interpreted [1-3] as phonon sidebands. This qualitative interpretation is quite relevant and attractive, however quantitative agreement between predictions for these phonon sidebands in frames of different models and the experiments is not yet achieved. Calculations of ARPES spectrum in t-J model [3,18] yield the band with the maximum at the binding energy 1.2 eV [3] that is much higher than in experiments (0.5-0.6 eV [1-6]). On the opposite, calculation of the optical conductivity band due to strong-coupling large polarons [19,20] yields the band with the maximum at approximately $E_p$. Experimental mid-IR bands in OC spectra of complex oxides have maximums at 0.5-0.8 eV [7-17]. However even at maximum for large polaron free-carrier effective mass $m^*$=2 (higher values of $m^*$ correspond ordinarily to the small polaron case) and typical effective dielectric permittivity $\varepsilon^*$ about 3-4 the SCLP binding energy [21] $E_p$=0.164-0.327 eV. (Effective dielectric permittivity

$$\varepsilon^* = \frac{\varepsilon_0 \varepsilon_\infty}{\varepsilon_0 - \varepsilon_\infty},$$ 

$\varepsilon_0$ is static dielectric constant, $\varepsilon_\infty$ is a dielectric permittivity at frequencies much higher than phonon ones)

Pekar [21] and Emin [22] considered an absorption spectrum caused by photodissociation of SCLPs. The maximum of the absorption band obtained by Pekar [21] and Emin [22] (at about $3E_p$) is in sufficiently good conformity with experimental OC spectra. (I.e. the polaron binding energy estimated as one third of the energy of maximum is in good conformity with that calculated according SCLP theory [21] with medium parameters typical for cuprates and other complex oxides.) However the form of the band predicted in [22] (strongly asymmetric with almost vertical low-energy edge) differs from the results of experiments.

We suggest common approach to calculate band in ARPES and OC spectra caused by photodissociation of strong-coupling large polaron (SCLP). Photodissociation is breaking of polaron due to photon absorption. Transverse electromagnetic field of photon does not interact with the longitudinal polarization field included in SCLP, it affects only the charge carrier. According to adiabatic condition satisfied in the SCLP case ($E_p/\hbar\omega \gg 1$ [21], where $\hbar\omega$ is the phonon energy) the photodissociation time (approximately $\hbar/(3E_p)$ [21,22]) is much shorter than characteristic phonon time $\omega^{-1}$. Therefore the phonon field does not have time to dress the charge carrier with a polarization cloud during the time of carrier phototransition and the state of the phonon field does not change during this time in conformity with Franck-Condon principle [19-22]. Thus, at the SCLP photodissociation the electron released by photon from the potential well appears in one of states of the continual energy spectrum. We will approximate the carrier states in continual spectrum as plane waves. In the case of photoemission when the photon energy is of the order of tens eV this approximation is ordinary. In the case of optical conductivity we can check the applicability of this approximation comparing the result of calculation with experiments. (We demonstrate below that the values of the polaron binding energy determined from experimental ARPES and OC spectra of the same material according to expressions obtained coincide with the error smaller than 10 %.)

Emin proposed such approach (approximation of final carrier states after SCLP photodissociation with plane waves) in calculation of optical absorption in work [22]. However, following Pekar [21], he used classical description for the polarization field in SCLP. In this approach the polarization field energy (and, hence, the number of radiated phonons) is fixed ($2E_p$). Accordingly, Pekar [21] predicted that optical dissociation of SCLP takes 3 times its binding energy in contrast with its thermal dissociation that takes only $E_p$. In conformity with this prediction the maximum of the band in the absorption spectrum calculated by Emin is situated at about $3E_p$ [22], the band has some half-width due to different possible values of the carrier momentum.

As distinct from Emin work [22] we consider the polarization field in SCLP quantum-mechanically [23,24] using the quantum-coherent states representation. It is shown in [23] that the SCLP includes the coherent lattice deformation. It is coordinated in phases superposition of the lattice states with different number of phonons so that one can call it phonon condensate. Since for the phonon part of SCLP its photodissociation is quick Franck-Condon process [19-22,24], the phonon condensate decays spontaneously (without participation of the electron-phonon interaction) when the charge carrier has left the potential well. Quantum-coherent state can be expanded in terms of states with the certain number of phonons [25,26], the expansion contains summands with all possible numbers of quanta. Therefore, decay of the coherent part of the polarization field at the SCLP photodissociation is accompanied by radiation of random number of phonons in each event. Only the average energy of phonons radiated in a single act is $2E_p$. As a result, corresponding band in OC spectrum is more wide and symmetrical than that calculated with classical consideration of the polarization field, and the position of its maximum ($4.2E_p$) differs from the predictions of classical theory ($3E_p$).

It should be noted that our consideration of ARPES spectrum does not take into account the electron losses at crossing the crystal. In OC spectrum we do not calculate the part of the spectrum caused by the charge carrier transitions into polaronic relaxed excited states (states of the discrete energy spectrum). This part of the OC spectrum of SCLP was calculated in works [19,20,27]. To obtain the complete OC spectrum of SCLP one should unite the band caused by SCLP photodissociation calculated here, the band caused by transitions into polaronic states and Drude contribution.

It is interesting to note that such complex structure is demonstrated by polaronic OC spectrum only in the case of strong electron-phonon coupling. In the case of weak and intermediate electron-phonon coupling the band caused by the polaron photodissociation is absent. The reason of this difference is the fact that separation of the polaron at the phototransition into two uncoupled parts can occur only in the case of strong electron-phonon coupling. In the weak and intermediate-coupling case Franck-Condon principle is not applicable so that the electron transition into a new state after the photon absorption is accompanied by a change of the lattice state. Indeed, in this case the time of the phototransition is of the same order as the characteristic phonon time $\omega^{-1}$. Therefore, because of the photon absorption the weak/intermediate-coupling polaron passes again into a polaron state (a state of the electron coupled with phonons, but not self-consistent at first [27]). I.e. the photodissociation of the polaron in the weak and intermediate coupling case cannot occur.

Thus, we suggest sufficiently simple and apparent method of calculating the band in ARPES and OC spectra caused by SCLP photodissociation. The method is based on using the coherent-state representation for the state of the phonon field in SCLP. Parameters of the coherent state of the polarization field as function of the medium parameters are obtained in [23] by variation. Then Fermi golden rule is applied to calculate the probability of the SCLP photodissociation with radiation of different number of phonons [24]. This probability allows calculating both ARPES and OC bands caused by SCLP photodissociation.

## 2. Band in ARPES and OC spectra caused by SCLP photodissociation

Let us briefly recall the calculation of the probability of the SCLP photodissociation with radiation of $\upsilon$ phonons [24] and then use it to obtain bands in ARPES and in OC spectrum caused by SCLP photodissociation. The probability $W_{if}$ of the system transition from the initial state $|i\rangle$ into the state $|f\rangle$ per unit time under the influence of the operator $\hat{H}_{int}$ according to Fermi golden rule is

$$W_{if} = \frac{2\pi}{\hbar} \left| \langle f | \hat{H}_{int} | i \rangle \right|^2 \delta(E_i - E_f), \quad (1)$$

where $E_i$ and $E_f$ are the energies of the initial and final states of the system. The polaron photodissociation occurs as a result of interaction of an electromagnetic wave of frequency $\Omega$ with the charge carrier in the polaron (the longitudinal field of the polarization in the polaron obviously does not interact with the transverse electromagnetic wave). The operator of the interaction has the form

$$\hat{H}_{int} = \frac{e\hbar(\mathbf{kA})}{m^*c} e^{i\mathbf{Q}\mathbf{r}}, \quad (2)$$

where $\hbar\mathbf{k}$ is the electron momentum, $\mathbf{A}$ is the amplitude of the vector potential of the electromagnetic field, related with its intensity $I$ as it follows: $I = \Omega\mathbf{A}^2/2\pi\hbar c$; $\mathbf{Q}$ is the wave vector of the electromagnetic wave.

At T=0K the initial state of the system is the ground state of the polaron in an electromagnetic field of a frequency $\Omega$. Therefore, the initial state and its energy are

$$|i\rangle = \sqrt{\beta^3/7\pi}(1+\beta r)\exp(-\beta r)\prod_{\mathbf{q}}|d_{\mathbf{q}}\rangle \quad (3)$$

and $E_i = -E_p + \hbar\Omega$, respectively, where Pekar wave function [21] for the carrier state in the polaron is used. The parameters $d_\mathbf{k} = |d_\mathbf{k}|e^{i\varphi_\mathbf{k}}$ of the coherent state of the phonon field in a ground state of SCLP found by variation in [23] have the form

$$|d_\mathbf{k}| = \frac{e}{|\mathbf{k}|}\sqrt{2\pi}(V\varepsilon^*\hbar\omega_\mathbf{k})^{-1/2}\eta_\mathbf{k}(\beta) \quad (4)$$

$$\varphi_\mathbf{k} = -\mathbf{k}\mathbf{R} + 2\pi C(\mathbf{k}\mathbf{R}), \quad (5)$$

where V is a crystal volume, $\omega_\mathbf{k}$ is the phonon frequency, $\eta_\mathbf{k}(\beta)$ is Fourier-transform of the square of modulus of the carrier wave-function in the polaron, β is the parameter characterizing the degree of the carrier localization in the polaron, **R** is the coordinate of the polaron center, $C(\mathbf{k}\mathbf{R})$ is an integer chosen in such a way that the phase $\varphi_\mathbf{k}$ belongs to the interval $[-\pi, +\pi]$.

Photodissociation of SCLP is Franck-Condon process since, according to adiabatic condition satisfied in the strong-coupling case, the polaron binding energy $E_p$ is much larger than the phonon energy $\hbar\omega$. Then the carrier phototransition time (in the case of strong-coupling large polaron it is about $\hbar/3E_p$ [21,22]) is much shorter than the characteristic phonon time $\omega^{-1}$. Thus, decay of the polaronic phonon "cloud" occurs without participation of electron-phonon interaction. The vectors of possible final states of the phonon field are a superposition of eigen vectors $|\{\nu_\mathbf{q}\}\rangle = \prod_\mathbf{q}|\nu_\mathbf{q}\rangle$ of the non-shifted Hamiltonian $\hat{H}_{ph} = \sum_\mathbf{q}\hbar\omega_\mathbf{q}b_\mathbf{q}^+b_\mathbf{q}$ describing the states with the certain number of quanta $\nu_\mathbf{q}$ in each harmonics. The electron final state after SCLP photodissociation as we discussed in the Introduction is free-carrier state not coupled with the polarization potential well and can be approximated as plane-wave state. Thus, after photodissociation the state (3) transforms into a state

$$|f\rangle = L^{-3/2}\exp(i\mathbf{k}\mathbf{r})\prod_\mathbf{q}|\{\nu_\mathbf{q}\}\rangle, \quad (6)$$

provided the sum of $\nu_\mathbf{q}$ (taking values 0 or 1) from the set $\{\nu_\mathbf{q}\}$ yields a certain number $\nu$. Hence, the energy of the final state is $E_f = \frac{\hbar^2\mathbf{k}^2}{2m^*} + \hbar\omega\nu$, if we neglect the dependence of $\omega$ on **q**, and

$$\delta(E_i - E_f) = \delta\left(-E_p + \hbar\Omega - \frac{\hbar^2\mathbf{k}^2}{2m^*} - \nu\hbar\omega\right). \quad (7)$$

The probability of the electron transition into a state with the wave vector with modulus $k$ and direction in a spatial angle $d\Omega$ around the direction determined by angles $\varphi, \theta$ has the form

$$dW_{\{\nu_\mathbf{q}\},\mathbf{k}} = \frac{2\pi}{\hbar}\left\{\frac{e\hbar(\mathbf{k}\mathbf{A})}{m^*c}32\sqrt{\frac{\pi}{7\beta^3}}L^{-3/2}\left(1+\beta^{-2}|\mathbf{Q}-\mathbf{k}|^2\right)^{-3}\right\}^2 \cdot \prod_\mathbf{q}|\langle\nu_\mathbf{q}|d_\mathbf{q}\rangle|^2 d\rho(\mathbf{k}), \quad (8)$$

where

$$d\rho(\mathbf{k}) = \frac{m^*L^3 k(\varepsilon)}{(2\pi)^3\hbar^2}d\Omega$$

is the spectral density of the final carrier states with the wave vector directed in the body angle $d\Omega$ [22]. According to Exp.(7) the electron momentum $\hbar\mathbf{k}$ and kinetic energy $\varepsilon$ in the final state are related as follows:

$$\hbar k(\varepsilon) = \sqrt{2m^*\varepsilon} = \sqrt{2m^*(\hbar\Omega - E_p - \nu\hbar\omega)}. \quad (9)$$

According to the energy and the momentum conservation laws (Exp.(7) and $\mathbf{Q} = \mathbf{k} + \mathbf{q}_0$, where $\mathbf{q}_0$ is the wave vector of the phonon field after the polaron photoionization, $\mathbf{q}_0 = \sum_{\mathbf{q}} \mathbf{q} \nu_{\mathbf{q}}$) an experiment can measure only the probability (8) summarized over all possible sets $\{\nu_{\mathbf{q}}\}$ having the same values of $\nu = \sum_{\mathbf{q}} \nu_{\mathbf{q}}$ and $\mathbf{q}_0$:

$$dW_{\{\nu_{\mathbf{q}}\},\mathbf{k}} = \frac{2\pi}{\hbar}\left\{\frac{e\hbar(\mathbf{kA})}{m^*c}32\sqrt{\frac{\pi}{7\beta^3}}L^{-3/2}\left(1+\beta^{-2}|\mathbf{Q}-\mathbf{k}|^2\right)^{-3}\right\}^2 \cdot \sum_{\{\nu_{\mathbf{q}}\}}^{\nu}\prod_{\mathbf{q}}|\langle\nu_{\mathbf{q}}|d_{\mathbf{q}}\rangle|^2 d\rho(\mathbf{k}). \quad (10)$$

Here symbol $\nu$ over $\Sigma$ denotes that the sum is carried out over the sets $\{\nu_{\mathbf{q}}\}$ satisfying the condition $\sum_{\mathbf{q}}\nu_{\mathbf{q}} = \nu$. Besides, there is not a set with $\nu = 0$ among the sets $\{\nu_{\mathbf{q}}\}$. Indeed, in such a case $\mathbf{q}_0 = 0$, hence, $\mathbf{k} = \mathbf{Q}$, and $\mathbf{kA} = \mathbf{QA} = 0$, i.e. the probability of a transition with appearance of such a set $\{\nu_{\mathbf{q}}\}$ is zero.

Sum in Exp.(10) expressing the probability of radiation of $\nu$ phonons in the SCLP photodissociation was calculated in [24]:

$$P_{\nu} = \sum_{\{\nu_{\mathbf{q}}\}}^{\nu}\prod_{\mathbf{q}}|\langle\nu_{\mathbf{q}}|d_{\mathbf{q}}\rangle|^2 = \frac{\overline{\nu}^{\nu-1}}{(\nu-1)!}e^{-\overline{\nu}} \quad (11)$$

where $\overline{\nu}$ is the average number of phonons radiated at the SCLP photodissociation [24]:

$$\overline{\nu} = 2E_p(\hbar\omega)^{-1}. \quad (12)$$

To calculate the band in ARPES caused by SCLP photodissociation we use ordinary for ARPES experiment geometry [1,2] where the wave vector $\mathbf{Q}$ of the incident photon lies in XZ plane of the coordinate system and makes angle ψ with z axes. XY plane of the coordinate system coincides with the sample surface. Wave vector $\mathbf{k}$ of the photoelectron inside the medium is considered to have a component $k_{||}$ lying in XY plane and $k_{\perp}$ in the perpendicular direction. The angle between $k_{||}$ and x-axis is φ. As ARPES measures the energy $\varepsilon'$ and momentum $\mathbf{k}'$ of the electron outside the medium we should express the probability (10) in terms of these variables. According to energy conservation law [1,2] the electron kinetic energy outside the medium

$$\varepsilon' = \frac{\hbar^2 k'^2}{2m_e} = \varepsilon - \Phi = \hbar\Omega - \Phi - E_p - \nu\hbar\omega, \quad (13)$$

where $\Phi$ is work function. The electron wave vector component lying in XY plane $k_{||}$ is continuous on the medium boundary [1,2]. The perpendicular component $k_{\perp}$ is discontinuous on the boundary [1,2]. If the dispersion of the electron inside and outside the medium differs only by its effective mass ($m^*$ and $m_e$, respectively) then

$$k_{\perp}' = \sqrt{(2m^*\varepsilon'/\hbar^2)(m^*/m_e) - k_{||}^2}$$

and

$$(\mathbf{kA}) = k_{||} \cos\varphi A\cos\psi + k_\perp' A\sin\psi \ . \qquad (14)$$

The wave vector **Q** of photons ordinarily used in ARPES experiments can be approximated as zero [1,2].

Thus, expression (10) takes the form

$$dW_{\{v_\mathbf{q}\},\mathbf{k}} = \frac{256 e^2}{7\pi\hbar m^* c^2 \beta^3} \frac{\left(k_{||}\cos\varphi\cos\psi + \sqrt{(2m^*\varepsilon/\hbar^2)(m^*/m_e) - k_{||}^2}\sin\psi\right)^2 A^2 k'}{(1+\beta^{-2}k'^2)^6} \cdot \frac{\overline{v}^{v-1}}{(v-1)!} e^{-\overline{v}} d\Omega$$
(15)

where the body angle $d\Omega$ contains a factor $\sqrt{2m^*\varepsilon'/\hbar^2 - k_{||}^2}/k_\perp'$ due to $k_\perp$ discontinuity on the medium boundary, $k'$ is determined by Exp.(13). Exp.(15) represents the probability of the polaron photodissociation at T=0K with appearance of $v$ phonons and electron with the kinetic energy $\varepsilon'$ and wave vector $\mathbf{k}'$ in the body angle $d\Omega$ around the direction determined by $\mathbf{k}'$ projections on X-, Y- and Z-axes $k_{||}\cos\varphi$, $k_{||}\sin\varphi$, and $k_\perp'$, respectively.

Band in OC spectrum caused by SCLP photodissociation can also be calculated on the base of Exp.(10). The real part of the conductivity has the form [28]

$$\mathrm{Re}\,\sigma = \frac{\hbar\Omega N_p \sum_v W(\Omega,v)}{\varepsilon_\infty E^2} \ . \qquad (16)$$

Here $W(\Omega,v)$ is probability (10) integrated over angular variables where the modulus of the carrier wave vector $k(\varepsilon)$ is expressed as function of photon energy $\hbar\Omega$ and of the radiated phonons number $v$ according to Exp.(9), $N_p$ is polaron concentration, and the fact that polarons interact with the light in a medium with a refraction index $\sqrt{\varepsilon_\infty}$ is taken into account. Using expression for $\beta$ in Pekar wave function [21] ($\beta = m^* e^2/(2\hbar^2 \varepsilon^*)$) one obtains

$$\mathrm{Re}\,\sigma = \frac{1024}{21} \frac{e^2 N_p}{m^*\Omega\sqrt{\varepsilon_\infty}} \sum_v \left[0.3\varepsilon(\Omega)\frac{\varepsilon^{*2}}{m^*/m_e}\right]^{3/2} \left[1 + 0.3\varepsilon(\Omega)\frac{\varepsilon^{*2}}{m^*/m_e}\right]^{-6} P_v,$$

or

$$\mathrm{Re}\,\sigma = \frac{1024}{21} \frac{e^2 N_p}{m^*\Omega\sqrt{\varepsilon_\infty}} \sum_v \left[\varepsilon(\Omega)\frac{0.44}{E_p}\right]^{3/2} \left[1 + \varepsilon(\Omega)\frac{0.44}{E_p}\right]^{-6} P_v \qquad (17)$$

where $\varepsilon(\Omega)$ is determined by Exp.(9).

3. Discussion of the results and comparison with other models and with experiments

Figs.1,2 demonstrate calculated by Exp.15 band in ARPES (the energy dispersion curve for $k_y=0$, $k_x=1$ in $\pi/a$ units where $a$ is the lattice constant) caused by photodissociation of SCLP with the binding energy 0.17 eV and 0.14 eV, respectively. For both figures the electron-phonon interaction constant α=6 (it determines the phonon energy for given $E_p$), $\hbar\Omega - \Phi$ =20eV and phonon dispersion is neglected. Abscissa axis is graduated in units of the so-called binding energy of electrons in the material [1,2], i.e. the difference between $\hbar\Omega - \Phi$ and the kinetic energy of photoemitted electrons.

As it is seen from Figs.1,2 the band calculated with neglected phonon dispersion consists of lines, each line correspond to certain number of radiated phonons. If do not neglect the phonon dispersion or if to take into account the finite lifetime ($\omega^{-1}$) of the charge carrier in a plane-wave (final) state the lines in Figs.1,2 will transform into bands, and the resulting summarized band can be structured or unstructured depending on the phonon dispersion (since a distance between neighbouring lines comprising the band is the phonon energy). As Figs.1,2 show the band in ARPES caused by SCLP photodissociation is wide. Its half-width (full width at half height) is in the interval 1.3 - 1.7$E_p$, depending on the phonon energy, or the value of α. (The half-width is 1.7$E_p$ for α=6 and is smaller for larger α.) The band maximum is situated approximately at the electron energy $\hbar\Omega - \Phi - 3.2E_p$, or at the binding energy 3.2$E_p$.

Fig.3 shows energy dispersion curves (their envelopes) calculated by Exp.(15) for $k_y=0$, $k_x$ changing from 0 up to 1 (in π/a units). As it is seen from Fig.3 the position of maximum in T=0K limit does not depend on the electron wave vector direction. This result is opposite to the case of small polarons calculated in t-J model [3, 18] where the band in ARPES caused by polarons demonstrates dispersion similar to the carrier dispersion in t-J model without phonons. As Fig.3 demonstrates, the intensity of the band in ARPES caused by SCLP photodissociation depends on the in-plane wave vector value in accordance with scalar product Exp.(14). Namely, the band intensity changes two times with the wave vector $k_x$ changing from 0 up to π/a at $k_y=0$. Approximately the same change of intensity is demonstrated by the band when the wave vector changes along the diagonal direction.

It is worth noting that intensity of the so-called "coherent" zero-phonon line (band) is equal to zero in T=0K limit. The reason of this as it was discussed above is following: if the number ν of radiated phonons is zero then the total wave vector of radiated phonons $\mathbf{q}_0 = 0$. Hence, according to the momentum conservation law in T=0K limit $\mathbf{k} = \mathbf{Q}$, and $\mathbf{kA} = \mathbf{QA} = 0$, i.e. the probability (10) of a transition with appearance of ν=0 phonons is zero in T=0K case.

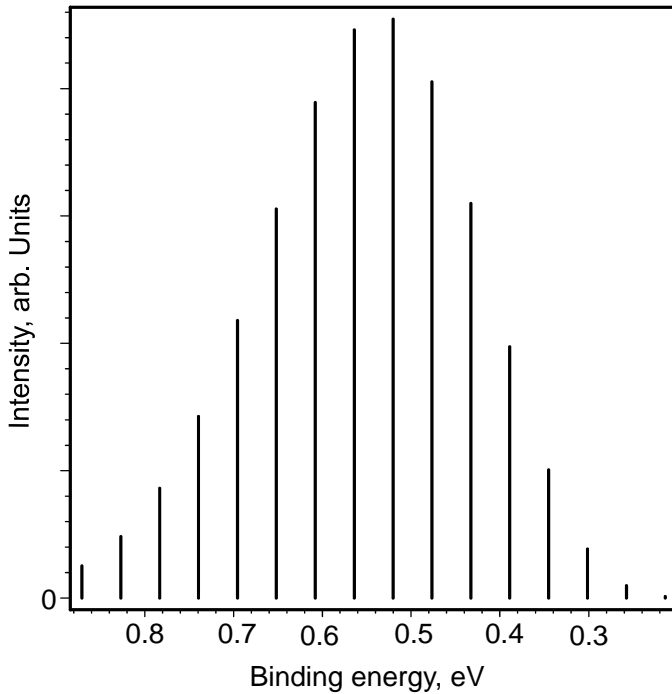

Fig.1. Band in ARPES caused by photodissociation of SCLPs with $E_p$=0.17 eV (energy dispersion curve at $k_x = \pi/a$, $k_y=0$) in neglect of phonon dispersion, at α=6 (that corresponds to the phonon energy $\hbar\omega$ =0.044 eV) and $\hbar\Omega - \Phi$ =20 eV.

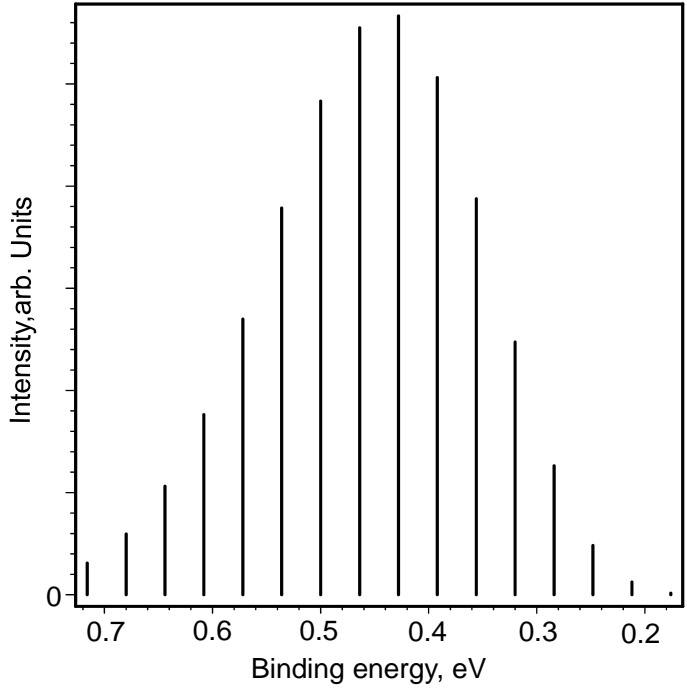

Fig.2. Band in ARPES caused by photodissociation of SCLPs with $E_p$=0.14 eV (energy dispersion curve at $k_x = \pi/a$, $k_y=0$) in neglect of phonon dispersion, at α=6 (that corresponds to the phonon energy $\hbar\omega$ =0.036 eV) and $\hbar\Omega - \Phi$ =20 eV.

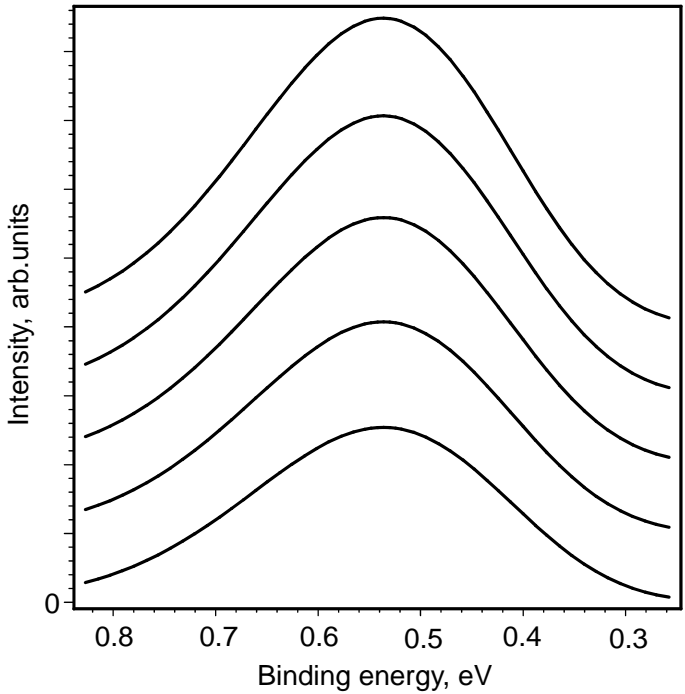

Fig.3. Energy dispersion curves (their envelopes) due to photodissociation of SCLPs with binding energy $E_p$=0.17 eV at α=6, $\hbar\Omega - \Phi$ =20 eV. Curves from the bottom to the top corresponds to $k_y$=0, $k_x$=0, 0.25, 0.5, 0.75, 1 in $\pi/a$ units, respectively.

We calculate the band in ARPES and OC spectra caused by SCLP photodissociation considering 3D medium. Nevertheless the result can be compared with experimental spectra in effectively 2D cuprates since the difference between 3D and 2D cases in such calculation is small as it was shown by Emin [22]. Comparison of the ARPES band due to SCLP photodissociation calculated according to Exp.(15) with experimental ARPES spectra of underdoped cuprates demonstrates good conformity. Experimental energy dispersion curves in ARPES spectra of underdoped cuprates contain broad bands with the maximum at binding energy $E_{max}$ shown in the second column of Table 1. The third column of Table 1 demonstrates the binding energies of SCLPs that can cause such bands, calculated as $E_{max}/3.2$ according to Exp.(15). The values of these binding energies and corresponding phonon energies are quite typical for strong-coupling large polarons in cuprates and other complex oxides.

Indeed, the adiabatic condition determining the strong-coupling case demands the polaron binding energy much higher than the phonon energy, the electron-phonon coupling constant α ≥ 6. It corresponds to average phonon number $\bar{v} = 2E_p(\hbar\omega)^{-1} \geq 7.77$. The phonon energy corresponding to these parameters $\hbar\omega = 2E_p/\bar{v} \leq E_p/3.89$. The values of phonon energy corresponding to the polaron binding energies from the third column of Table 1 are in good conformity with experimental data for cuprates. For example, the phonon branches strongly interacting with the charge carrier in $Nd_{1.85}Ce_{0.15}CuO_4$ have the frequency 115, 130, 146, 220 and 305 cm$^{-1}$ [29]. The energy of phonons strongly interacting with the charge carrier in copper-containing complex oxides were also estimated on the base of experimental data in [30] as the value of the order of 0.02 eV.

Table 1. Position of maximum $E_{max}$ of band in ARPES observed experimentally in different materials and polaron binding energies $E_p$ calculated from them.

| Material | $E_{max}$, eV | $E_p$, eV |
|---|---|---|
| $La_{2-x}Sr_xCuO_4$ (x=0,0.03) | ≈0.48 [4] | ≈0.15 |
| $Nd_{2-x}Ce_xCuO_4$ (x=0.04) | ≈0.39 [5] | ≈0.122 |
| $Ca_{2-x}Na_xCuO_2Cl_2$ (x=0,0.05) | ≈0.5 [6] | ≈0.156 |
| $La_2CuO_{4+y}$ | About 0.5 [3] | ≈0.156 |

The form of the bands calculated according to Exp.(15) is also in good conformity with the experiments. The half-width of the experimental band estimated from its lower binding energy part as in [3] and half-width of the band calculated according to Exp.(15) are close. However the experimentally measured edge of the band from the higher binding energy side is enhanced essentially in comparison with the lower binding energy part [3]. Different mechanisms of carries scattering at crossing the crystal that we do not take into account in our calculation can be responsible for this effect.

As we use the common approach to calculate bands due to SCLPs photodissociation in ARPES and OC spectrum we obtain them in terms of the same parameters – polaron binding energy $E_p$ and phonon energy. Therefore one can predict the position of maximum and the form of the band in ARPES on the base of the maximum position of OC mid-IR band and vice versa. To do it let us consider the band in OC spectrum caused by SCLP photodissociation determined by Exp.(17).

An example of these bands is shown by Fig.4 for α=6 (curve 1) and α=8 (curve 2) that corresponds to $E_p$ = 0.117 eV and $E_p$ = 0.207 eV, respectively, if the phonon energy $\hbar\omega$ =0.03 eV. More precisely, Fig.4 demonstrates $\sigma(\Omega)/\left[\sqrt{\varepsilon_\infty}(m^*/m_e)\right]$ for a system with the polaron concentration $N_p = 10^{18} cm^{-3}$. Since Exp.(17) represents the optical conductivity in linear in the polaron number approximation, one can easily obtain the spectra for any other polaron concentration. As it is seen from Fig.4 the absorption band due to SCLP photodissociation turns out to be unstructured. It is natural as the band consists of "partial" bands, each "partial" band corresponds to certain value of number ν of radiated phonons. The distance between neighbor "partial" bands is $\hbar\omega$ whereas the width of each "partial" band is determined by a condition *kR<1* where *R* is the polaron radius [22] (while the condition is satisfied the matrix element is significant) and is of the order of $E_p$ [22]. According to adiabatic condition satisfied in the strong-coupling case $E_p >> \hbar\omega$, hence, the band is unstructured.

As Fig.4 demonstrates the optical conductivity band caused by the SCLPs photodissociation is wide band with a maximum at $\Omega_{max} \approx 4.1 \div 4.2 E_p/\hbar$, a half-width $\Delta\Omega \approx 2.2 \div 2.8 E_p/\hbar$ (in case of α=6-8) and a long-wavelength edge $\Omega_{edge} = E_p/\hbar + \omega$, as $\nu \geq 1$. Increase in the polaron binding energy results in an increase of $\Omega_{max}$ and $\Delta\Omega$ (curve 2). However, being expressed in units of $E_p/\hbar$ the value $\Omega_{max}$ remains unchanged: $\Omega_{max} \approx 4.2 E_p/\hbar$ whereas $\Delta\Omega$ even decreases from 2.8 $E_p/\hbar$ to 2.2 $E_p/\hbar$ at the increase of α. Thus, in the case of α=6-8 one can calculate the polaron binding energy from its absorption spectrum as $E_p = \hbar\Omega_{max}/4.2$.

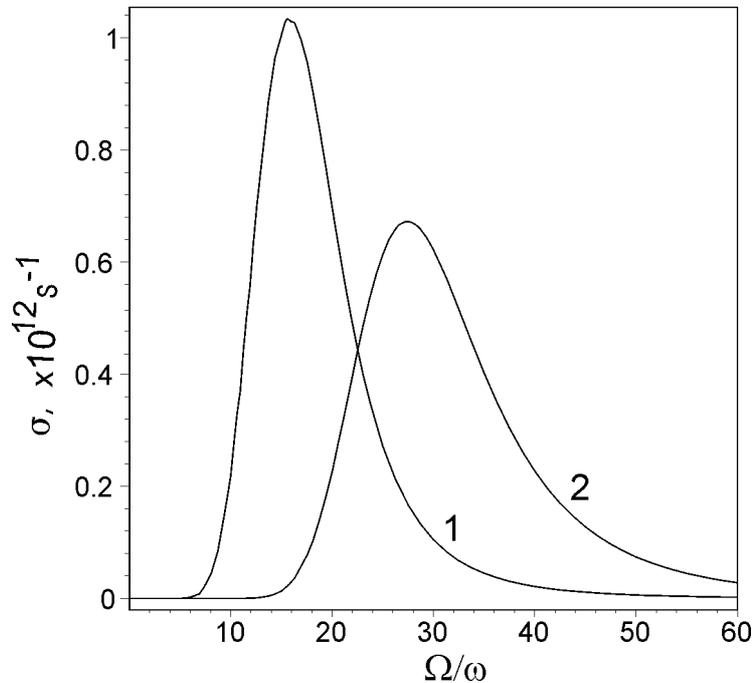

Fig.4. Optical conductivity $\sigma/(\sqrt{\varepsilon_\infty} m^*/m_e)$ caused by photodissociation of Landau-Pekar polarons at their concentration $N_p = 10^{18} cm^{-3}$. Curve 1 corresponds to α=6 ($\bar{v} = 7.776$, $E_p$ =0.1125 eV if the phonon energy $\hbar\omega$ =0.03 eV). Curve 2 corresponds to α=8 ($\bar{v} = 13.824$, $E_p$ =0.207 eV if $\hbar\omega$ =0.03 eV).

It is useful to compare the band in OC spectrum due to SCLP photodisociation calculated at classical [22] and at quantum-mechanical consideration of the polarization field. Curve 1 on Fig.5 is calculated according to Emin expression [22], curve 2 is calculated by Exp. (17) but with the carrier wave function in the polaron used by Emin, curve 3 is calculated by Exp.(17) with Pekar wave function [21] of the carrier in the polaron. As it is seen from Fig.5 the spectrum obtained in the present work differs essentially from that predicted by Emin [22]. They differ in the low-frequency edge position: $\hbar\Omega_{edge} = 3E_p$ in [22] and $\hbar\Omega_{edge} = E_p + \hbar\omega$ in the present work; in the band maximum position: $\hbar\Omega_{max} = 3 \div 3.5 E_p$ in [22] and $\hbar\Omega_{max} = 4.2 E_p$ in the present work; and in the half-bandwidth of the band: $\hbar\Delta\Omega \approx 1.5 E_p$ in [22] and $\hbar\Delta\Omega = 2.2 \div 2.8 E_p$ in the present work. The band obtained in the present work is more symmetrical with a stretched low-frequency tail apart from the high-frequency one.

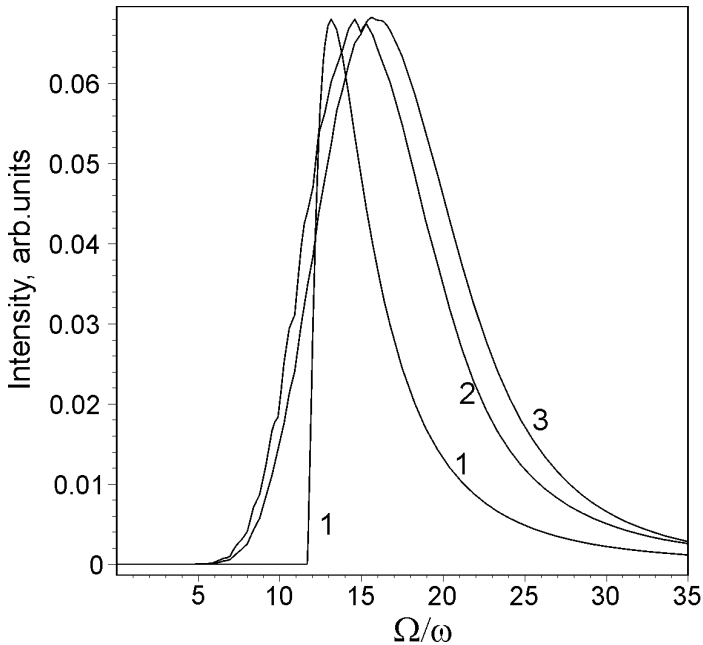

Fig.5. Light absorption caused by photodissociation of SCLPs. Curve 1 is calculated according to [22], curve 2 and curve 3 are calculated by Exp.(17) with Emin [22] and Pekar [21] wave function of the charge carrier in the polaron, respectively. For all the curves $\bar{v} = 7.776$ (i.e. α=6), $E_p$ =0.1125 eV, the phonon energy $\hbar\omega$ =0.03 eV.

Thus, the OC band caused by the SCLP photodissociation is wide unstructured band with a single maximum at the frequency $\Omega_{max} \cong 4.2 E_p/\hbar$ and a half-bandwidth of the order of $2.2 \div 2.8 E_p/\hbar$ (in the case of α=6-8). It is natural to compare this band with mid-infrared bands observed in OC spectra of non-stoichiometric (doped to obtain excess charge carriers) complex oxides: cuprates [7-16] and nikelates [17]. The maximums of mid-IR bands in OC spectra of complex oxides are shown in the second column of Table 2. The values of $E_p$ calculated for these substances from the band maximum positions $\hbar\Omega_{max}$ according to the result of the present consideration as $\hbar\Omega_{max}/4.2$ are shown in the third column of Table 2. They are in good conformity with $E_p$ values calculated according to SCLP theory [21] with the use of typical for complex oxides medium parameters $\varepsilon^* \approx 3-4$, $m^*/m_e$=1-2 ($E_p$=0.092-0.327eV). Obviously, interpretation of these bands as bands caused by carrier transitions into polaronic relaxed excited states calculated in works [19,20] where the energy of the band maximum is about $E_p$ ($E_p \approx \hbar\Omega_{max}$) results in the polaron binding energies too large for the large radius polarons. Interpretation of these experimental mid-IR bands as caused by SCLPs photodissociation does not lead to such contradiction.

However, the bands caused by phototransitions into polaronic relaxed excited states calculated in Refs.19,20 can also be found in OC spectra of complex oxides. Several groups reported the experimental spectra pointed out that the mid-IR bands in OC spectra of complex oxides have complex structure in which two bands can be distinguished [14-16]. Column 4 of Table 2 shows the maximums of the low-frequency mid-IR bands in experimental OC spectra. If we suppose that these bands are caused by transitions into polaronic relaxed excited states studied in [27,19,20] then according to [19,20] their maximums are located at approximately $E_p$. Comparison of the polaron binding energies calculated from the maximum position of high-frequency mid-IR bands (column 3) with the energies of maximum of the low-frequency mid-IR bands (column 4) demonstrates good agreement of our supposition with the experiment. This can also be seen from the comparison of the ratio $\Omega_{max}^{photodiss}/\Omega_{max}^{internal}$ obtained from experimental OC spectra [14-16] ($\Omega_{max}^{photodiss}/\Omega_{max}^{internal}$ = 4 - 4.5) with that predicted by united results of the present work and those of works [19,20]: $\Omega_{max}^{photodiss}/\Omega_{max}^{internal} \cong 4.2$.

Table 2. Maximum position of mid-IR bands in OC spectrum observed experimentally ($\hbar\Omega_{max}^{photodiss}$ is the maximum of high-frequency mid-IR band and $\hbar\Omega_{max}^{internal}$ is the maximum of low-frequency mid-IR band) and the polaron binding energy $E_p$ calculated from $\hbar\Omega_{max}^{photodiss}$.

| Material | $\hbar\Omega_{max}^{photodiss}$, eV | $E_p$, eV | $\hbar\Omega_{max}^{internal}$, eV |
|---|---|---|---|
| Yba$_2$Cu$_3$O$_{6+y}$ | 0.62±0.05 [14] | ≈0.155±0.01 | 0.16±0.03 [14] |
| Nd$_2$CuO$_{4-y}$ | 0.76±0.01 [14] | ≈0.18 | 0.162±0.005 [14] |
| La$_{2-x}$Sr$_x$CuO$_{4+y}$ | 0.53±0.05 [14] | ≈0.126±0.01 | 0.16±0.03 [14] |
| La$_2$CuO$_{4+y}$ | 0.6±0.02 [14] | ≈0.143±0.005 | 0.13±0.02 [14] |
| Nd$_{2-x}$Ce$_x$CuO$_4$ (x=0.05) | 0.55 [15] | ≈0.131 | 0.11 [15] |
| Nd$_{2-x}$Ce$_x$CuO$_4$ (x=0.1) | 0.39 [15] | ≈0.093 | 0.09 [15] |

Thus, presence of two bands in the OC spectrum of complex oxides is in good conformity with the predictions of the SCLP theory. The form of the high-frequency mid-IR bands in complex oxides is also in good conformity with that calculated by Exp.(17). It is demonstrated by Fig.6 where we fit the experimental OC spectrum [14] with the use of Exp.(17). The width of the band calculated theoretically would be larger if we took into account the limitation of the wave vector space in the First Brillouin zone. Besides, we calculate the band for the case of medium with a single phonon branch strongly interacting with the charge carrier whereas in complex oxides there are several such branches ordinarily. Taking into account this fact will lead to a variety of the band forms. Allowing for interaction of polarons will likely also lead to broadening of the band as it occurs in the weak-coupling case [31].

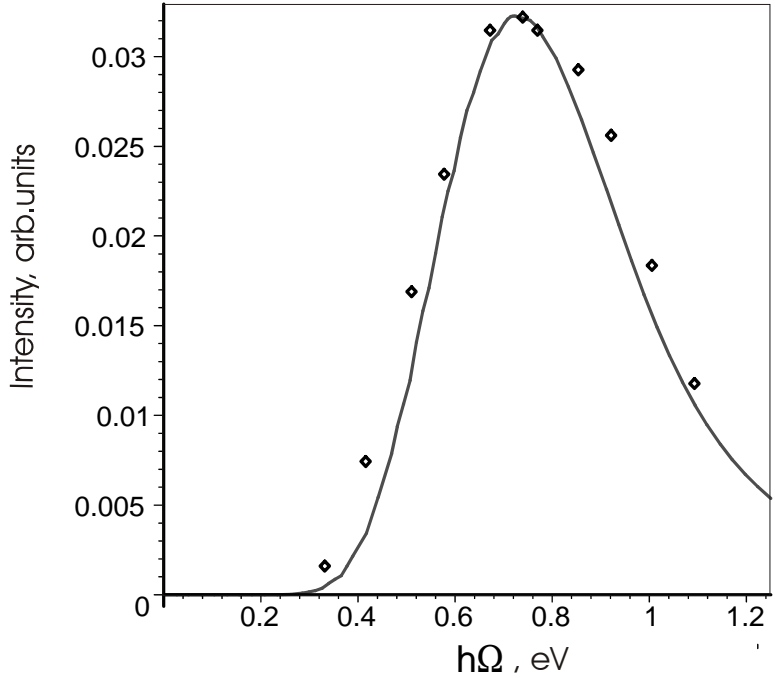

Fig.6. Mid-infrared band in the optical conductivity spectrum of $Nd_2CuO_{4-y}$ [14] at temperature T=10K (diamonds) and its fit by Exp.(17) with $\alpha=6$, $E_p=0.18$ eV (solid curve).

Now let us compare the polaron binding energies calculated from experimentally observed ARPES band (as $E_{max}/3.2$) and from mid-IR band in OC spectrum (as $\hbar\Omega_{max}/4.2$). They are presented in Table 3. Unfortunately we are not aware of both ARPES and OC spectra for a material with precisely the same doping level, only for somewhat different doping levels. This can lead to worse coincidence since the polaron binding energy and the maximum position of the bands depends on the doping level, as it will be discussed in the next section. Nevertheless, as Table 3 shows the polaron binding energies $E_p$ calculated from ARPES and OC spectra for the same material with close doping level demonstrate good agreement (the error is within 10 %). Hence, the approximation of final carrier states with plane waves used in calculation of OC band caused by SCLP photodissociation as it is discussed in Introduction dos not lead to large error and can be applied.

Table 3. Comparison of binding energies of SCLP estimated from ARPES and OC spectra.

| Material with measured ARPES | $E_p$ calculated from ARPES | Material with measured OC spectrum | $E_p$ calculated from OC spectrum |
|---|---|---|---|
| $Nd_{2-x}Ce_xCuO_4$ (x=0.04) | ≈0.122 eV [5] | $Nd_{2-x}Ce_xCuO_4$ (x=0.05) | ≈0.131 eV [15] |
| $La_{2-x}Sr_xCuO_4$ (x=0,0.03) | ≈0.15 eV [4] | $La_{2-x}Sr_xCuO_{4+y}$ | ≈0.126±0.01 eV [14] |
| $La_2CuO_{4+y}$ | ≈0.156 eV [3] | $La_2CuO_{4+y}$ | ≈0.143±0.005 eV [14] |

Thus, the result of the present consideration allows predicting the position of maximum of band in ARPES on the base of maximum position of mid-IR band in OC spectrum (using a factor 3.2/4.2) and vice versa. An example of such prediction is shown by Table 4. The second column of Table 4 contains the maximums of mid-IR bands obtained experimentally in works [7-13] and collected together in [32]. The third column of Table 4 demonstrates the binding energy of SCLPs calculated from the maximum positions. The fourth column of Table 4 shows the predicted positions of maximum of ARPES bands (the binding energy in eV) in these materials.

Table 4. Experimentally observed maximums of mid-IR band in OC spectrum of different materials and predicted position of maximum of band in ARPES.

| Material | Maximum of mid-IR band in OC spectrum, eV | Binding energy $E_p$ of the polaron, eV | Predicted position of maximum in ARPES, eV |
|---|---|---|---|
| $Bi_2Sr_2CaCu_2O_8$, a axis | 0.545 [7] | 0.129 | 0.415 |
| $Bi_2Sr_2CaCu_2O_8$, b axis | 0.51 [7] | 0.121 | 0.388 |
| $La_{1.9}Ca_{1.1}Cu_2O_{6+\delta}$ | 0.59 [8] | 0.14 | 0.45 |
| $La_{1.85}Sr_{0.15}CaCu_2O_{6+\delta}$ | 0.57 [8] | 0.136 | 0.43 |
| $La_{1.9}Sr_{0.1}CuO_{4+\delta}$ | 0.46 [9] | 0.109 | 0.35 |
| $YBa_2Cu_4O_8$ | 0.55 [10] | 0.13 | 0.42 |
| $HgBa_2CuO_{4+\delta}$ | 0.48 [11] | 0.114 | 0.366 |
| $Bi_2Sr_2Ca_2Cu_3O_{10}$ | 0.63 [12] | 0.15 | 0.48 |
| $YBa_2Cu_4O_{6.9}$ | 0.65 [13] | 0.155 | 0.495 |

4. Influence of temperature and carrier concentration on ARPES and OC spectra caused by SCLPs

We calculate the OC and ARPES bands caused by the SCLP photodissociation in T=0K approximation, i.e. for resting or moving slowly (with the velocities considerably lower than the maximum group velocity of phonons interacting with the carrier) Landau-Pekar polarons. Influence of the temperature on the bands will naturally lead to an increase of their width. However, the theory of photodissociation of Landau-Pekar polarons in case of non-zero temperatures is complicated essentially. Coherent polarization in Landau-Pekar polaron cannot propagate with a velocity higher than the group velocity of phonons participating in the polaron formation [23,33]. In the case considered by Pekar and in the present work we use the model with dispersionless phonons that corresponds to zero velocity of phonons. In such a model one cannot consider thermal motion of polarons. Besides, the distribution function of polarons has peculiarities caused by a competition of polaron and Bloch states of a charge carrier [34,35]. The problem of ARPES and OC spectra of polarons at finite temperatures will be considered elsewhere.

However, we can predict the change of the integral intensity of bands in ARPES and OC spectra caused by the SCLP photodissociation with the temperature. The characteristic feature of the SCLP is the limitation of its velocity by the maximum group velocity of phonons participating in its formation [33]. It results in thermal destruction of the SCLPs when their thermal velocities exceed the minimum phase velocity of relevant phonons. The thermal destruction expresses itself in a gradual decrease of the polaron concentration with temperature at critical temperatures much lower than $E_p$. (The critical temperature is determined by the values of the phonon maximum group velocity and the polaron binding energy [34,35].) Therefore, the integral intensity of the band will decrease with the increase of temperature.

An example of the polaron thermal destruction at temperatures low in comparison with the polaron binding energy is represented by optical conductivity spectrum of $\beta - Na_{0.33}V_2O_5$ [36]. It contains a mid-IR band with the maximum at 3000 $cm^{-1}$ that can be caused by photodissociation of SCLP with the binding energy about 0.09 eV. (Indeed, the form of the band in the low-temperature spectrum (T=5K) is in good conformity with its fit by Exp.(14) with $E_p$ =0.089 eV, α=6. The phonon energy corresponding to these parameters ($\hbar\omega = 2E_p/\overline{\nu}$ = 0.025 eV) is in good conformity with the reflectivity spectra [36], demonstrating large longitudinal-transverse splitting (i.e. strong electron-phonon interaction) for a set of phonon branches with the frequencies 80-300 cm$^{-1}$). Already the optical conductivity spectrum of $\beta - Na_{0.33}V_2O_5$ corresponding to T=145K demonstrates decrease of the integral intensity of the mid-IR band, and the 300K spectrum shows approximately two times lower integral intensity in comparison with that for T=5K spectrum [36]. Similar decrease of the integral intensity of mid-IR band in OC spectrum is demonstrated by $Nd_{2-x}Ce_xCuO_4$ [15].

Transition of the charge carriers from the polaron states into the free carrier states with increasing temperature results also in essential lowering their effective mass. Therefore, some decrease of the system resistance (the so-called activational behavior) may occur in the corresponding temperature interval [35] if the change of the carrier relaxation time at the transition will not compensate this effect. Some complex oxides demonstrate such a behavior of the system resistance and Hall coefficient [15,37]. For example, the integral intensity of mid-IR band in OC spectrum of $LaTiO_{3.41}$ decreases 3-4 times in the region of temperatures 150-300 K [37]. In the same temperature region the resistance along the direction ∥a demonstrates change of its temperature behavior: in the region 50-150 K the resistance increases with the temperature increase whereas in the region 150-300 K the increase stops and the resistance remains practically constant with increasing temperature. It is in conformity with the temperature behavior of resistance in systems with Landau-Pekar polarons [35].

We calculate the bands in ARPES and OC spectra in the limit of non-interacting polarons so that our results are valid only in the case of low carrier concentration (small doping level). However on the base of carrier distribution function in systems with strong-coupling polarons [34] we can predict some consequences of increase of the carrier concentration. According to the carrier distribution function in systems where SCLP can form [34], free charge carriers coexist with polarons at non-zero temperatures or at carrier concentration higher than the maximum polaron concentration $n_0=2V_0^{-1}$, where $V_0$ is the polaron volume. Therefore already at doping level higher than approximately 0.03 one can expect in ARPES features caused by free charge carriers together with those caused by polarons. Besides, screening of the electron-phonon interaction by free charge carriers will cause lowering the polaron binding energy and, consequently, decrease of the energy corresponding to the mid-IR band maximum at increase of doping as it is demonstrated by experiment [15].

5. Conclusion

Thus, according to the result of the present consideration the band in ARPES caused by SCLP photodissociation is wide non-dispersing band with the maximum approximately at the binding energy $3.2E_p$ and the half-width about $1.5E_p$. According to the carrier distribution function in system with SCLPs [33] free charge carriers coexist with polarons already at the doping level higher than 0.03. One can expect corresponding features in ARPES together with the band caused by SCLP photodissociation. In OC spectrum SCLP photodissociation results in the band with the maximum at approximately $4.2E_p$ and the half-width about $2.5E_p$. However together with it and Drude contribution OC spectrum of SCLPs contains another band caused by transitions into polaronic relaxed excited states (states of discrete energy spectrum). According to [19,20] the maximum of this band is situated at approximately $E_p$. Bands in ARPES and OC spectra caused by SCLP photodissociation are wide in conformity with experiment as the method suggested allows to describe appearance of many phonons in a single act of SCLP photodissociation (different number of phonons in different acts).

The common approach used to calculate ARPES and OC bands caused by SCLP photodissociation allows coordinated interpretation of ARPES and OC spectra of complex oxides. Using the results of the present consideration one can predict the position of the ARPES band maximum and half-width from the position of maximum of mid-IR band in OC spectrum and vice versa. Comparing the results of the present consideration with experiments we can conclude that underdoped cuprates contain SCLPs. The values of the polaron binding energy calculated according to Exps.(15,17) from ARPES and OC spectra of the same material are in good conformity: the difference between them is within 10 %. This confirms the applicability of the supposition made that plane-wave states can be used for approximation of final carrier states in calculation of OC band caused by SCLP photodissociation.

Acknowledgement. We are grateful to O. Gunnarsson, A. A. Kordyuk, A.V. Boris, M. Lindroos, and Ph. Aebi for helpful discussions.